\documentclass[preprint,12pt]{elsarticle}

\usepackage{amssymb}

\usepackage{amsmath}

\journal{Signal Processing}

\usepackage{color,xcolor}

\usepackage{amsthm}

\newtheorem{defn}{Definition}

\usepackage{algorithmic}
\usepackage{algorithm}

\usepackage{booktabs}
\usepackage{multirow}

\usepackage{epstopdf}

\usepackage[caption=false,font=normalsize,labelfont=sf,textfont=sf]{subfig}

\begin{document}

\begin{frontmatter}

\title{A Unified Algorithmic Framework for Dynamic Compressive Sensing}

\author[1]{Xiaozhi Liu}
\ead{xzliu@buaa.edu.cn}

\author[1]{Yong Xia \corref{cor1}}
\ead{yxia@buaa.edu.cn}

\cortext[cor1]{Corresponding author}

\affiliation[1]{organization={School of Mathematical Sciences},
            addressline={Beihang University}, 
            city={Beijing},
            postcode={100191},
            country={China}}

\begin{abstract}

We present a unified algorithmic framework, termed PLAY-CS, for dynamic tracking and reconstruction of signal sequences exhibiting intrinsic structured dynamic sparsity. By leveraging specific statistical assumptions on the dynamic filtering of these sequences, our framework integrates a variety of existing dynamic compressive sensing (DCS) algorithms. This is facilitated by the introduction of a novel Partial-Laplacian filtering sparsity model, which is designed to capture more complex dynamic sparsity patterns. Within this unified DCS framework, we derive a new algorithm, $\text{PLAY}^{+}$-CS. Notably, the $\text{PLAY}^{+}$-CS algorithm eliminates the need for a priori knowledge of dynamic signal parameters, as these are adaptively learned through an expectation-maximization framework. Moreover, we extend the $\text{PLAY}^{+}$-CS algorithm to tackle the dynamic joint sparse signal reconstruction problem involving multiple measurement vectors. The proposed framework demonstrates superior performance in practical applications, such as real-time massive multiple-input multiple-output (MIMO) communication for dynamic channel tracking and background subtraction from online compressive measurements, outperforming existing DCS algorithms.
\end{abstract}


\end{frontmatter}

\section{Introduction}
In numerous real-time applications, particularly those involving precise dynamic tracking of Millimeter Wave (mmWave) Channel State Information (CSI) in a massive Multiple-Input and Multiple-Output (MIMO) system employing a Hybrid Analog and Digital Beamforming (HBF) architecture \cite{mimo} and simultaneously background extraction in video frames \cite{BSCM_2016}, a fundamental challenge emerges. This challenge involves the causal reconstruction of time-varying signal sequences from limited compressive measurements.

Such causal signal reconstruction is sometimes referred to as dynamic filtering \cite{rwl1-df}. The goal of dynamic filtering is to solve the causal minimum mean squared error (MMSE) estimation of the {time-varying} signals. The conventional MMSE solution is embodied by the Kalman filter (KF) \cite{kf}. 
However, the KF assumes Gaussianity and linearity, conditions that are not always met in practice, particularly when the statistical properties of signals and innovations deviate from Gaussian distributions.
Moreover, the KF fails to harness the inherent dynamic structured sparsity of the signal sequences. The problem of recovering sparse time-varying signal sequences from a small number of compressive measurements has been extensively studied, often under the terms dynamic sparse recovery or dynamic compressive sensing (DCS) \cite{dcs-review}.

Most existing DCS solutions are based on batch algorithms \cite{batch-1,batch-2}. However, these offline methods are computationally prohibitive for massive MIMO systems due to their excessive memory requirements and limited scalability. Moreover, batch algorithms typically assume that the support set of sparse signals remains static over time, an assumption often inconsistent with real-world scenarios. This highlights the critical need for recursive algorithms capable of addressing these challenges.

Recursive algorithms offer significant advantages over batch techniques by reducing computational and storage complexity. They are often employed to integrate the observation model and prior information within a unified framework. A series of DCS methods have been proposed to address these issues \cite{kfcs,lscs,modcs,regmodbpdn,rwl1-df}, and are summarized as follows.

\textbf{Algorithms Exploiting the Slow Support Change of Dynamic Signals:} 
The Least Squares Compressive Sensing (LS-CS) algorithm \cite{lscs} is among the earliest methods to leverage the slow support change characteristic of signal sequences.
It begins by computing an initial LS estimate of the signal on the estimated support set and then performs $\ell_1$-norm minimization on the signal residual.
However, the performance of LS-CS is heavily dependent on accurate support set estimation, which can be challenging in practical applications.
The Modified-CS algorithm \cite{modcs} reframes the sparse recovery problem as one of finding the sparsest solution outside the support set while satisfying the measurement constraint.
Modified-CS can exactly recover the signal with fewer independent columns of the measurement matrix than the Regular-CS algorithm \cite{bp}, i.e., a {Compressive Sensing (CS)} algorithm that reconstructs each sparse signal in the sequence independently without leveraging prior information.
Furthermore, Modified-CS is a special case of Weighted-$\ell_1$ \cite{rwl1-df}, where the index set is divided into distinct subsets with different weights applied to the $\ell_1$ norm within each subset. 
The reweighted $\ell_1$ dynamic filtering (RWL1-DF) algorithm \cite{rwl1-df} based on the dynamic filter, employs a similar reweighting scheme in \cite{rwl1} and further improves the recovery performance.

\textbf{Algorithms Exploiting the Slow Value Change of Dynamic Signals:}
Mota \textit{et al.} \cite{Mota_2017_TIT,Mota_2016_TSP} demonstrated that solving an $\ell_1$-$\ell_1$ minimization problem can significantly enhance the performance of CS \cite{bp,bp2} when high-quality prior information is available. 
They also {employed} a motion-compensated extrapolation technique \cite{Mota_2016_TSP} to obtain accurate prior prediction.
However, the performance of the $\ell_1$-$\ell_1$ minimization method may degrade when there is poor correlation between the prior and target signals.
To address this issue, Luong \textit{et al.} proposed the Reconstruction Algorithm with Multiple Side Information using Adaptive Weights (RAMSIA) \cite{Luong_2016_ICIP} and the Compressive Online Robust Principal Component Analysis (CORPCA) model \cite{Luong_2018_TIP}, which balance the contributions from multiple prior information signals.

\textbf{Algorithms Exploiting the Slow Support and Value Change of Dynamic Signals:} 
The Kalman-filter type method, KF-CS, introduced in \cite{kfcs}, integrates a Kalman filter process within the support set while performing additional detection on filtering errors outside the support.
An extension of the Modified-CS algorithm, known as Reg-mod-BPDN, was proposed in \cite{regmodbpdn} to further exploit the slow-changing characteristic of the signal values. This method adds regularization to the optimization objective, enhancing the utilization of prior information on signal values.

Despite their utility, these methods do not fully exploit the structured dynamic sparsity inherent in signal sequences. While previous studies, such as \cite{GSM-zhanglei, LSM-nips2010}, have explored the hierarchical structured sparsity of the signals, they fall short of leveraging dynamic structured sparsity, which is crucial for reconstructing dynamic signal sequences. This highlights the necessity for developing a more advanced dynamic signal model and a more efficient algorithmic framework to better reconstruct dynamically structured sparse signal sequences.

Furthermore, in the expansive literature on DCS analysis, it is not yet fully understood whether these existing recursive DCS algorithms have some intrinsical correlations or not.
In this paper, we address this gap by introducing a structured sparsity model based on the dynamic filter. We construct a unified DCS algorithmic framework that not only provides deeper insights into existing DCS algorithms but also leads to the development of superior variants.

The main contributions of this paper are summarized as follows:
\begin{itemize}
	\item \textbf{Partial-Laplacian Filtering Sparsity Model:} 
	We propose a novel Partial-Laplacian filtering sparsity model designed to capture the structured sparsity inherent in dynamic signal sequences. This model is flexible and effective in capturing the dynamic filtering characteristics of practical signal sequences.

	\item \textbf{Unified DCS Algorithmic Framework:}
	By integrating the Kalman filter methodology with the proposed Partial-Laplacian model, we develop a unified DCS algorithmic framework, termed Partial Laplacian Dynamic CS (PLAY-CS). PLAY-CS reveals the correlations among existing DCS algorithms, demonstrating that several existing algorithms can be derived as special cases under specific conditions. This unification provides a comprehensive understanding of these methods, which is the rationale behind our chosen title.
	
	\item \textbf{Design of Specific Algorithms for Practical Applications: }
	Our framework enables the derivation of various DCS algorithms based on the proposed Partial Laplacian Scale Mixture (Partial-LSM) sparsity model. Among these, we introduce the novel $\text{PLAY}^{+}$-CS algorithm, which autonomously learns dynamic signal parameters using the Expectation-Maximization (EM) framework, eliminating the need for prior knowledge. Furthermore, $\text{PLAY}^{+}$-CS is extended for channel reconstruction in massive MIMO orthogonal frequency division multiplexing (OFDM) systems by exploiting dynamic joint sparsity across the angle-frequency-time domains, resulting in the $\text{PLAY}^{+}$-CS-MMV algorithm. Extensive experiments demonstrate that the proposed algorithms outperform existing DCS algorithms in practical applications, including dynamic channel reconstruction and Background Subtraction from Compressive Measurements (BSCM).
\end{itemize}

\textit{Outline}: In Section \ref{sec:problem}, we present the problem definition and the system model.
Section \ref{sec:application} reviews the applications of the DCS problem. 
In Section \ref{sec:algorithm}, we introduce the proposed Partial-Laplacian model along with the unified DCS algorithmic framework: {PLAY-CS}.
Furthermore, we derive a more efficient DCS algorithm, denoted as $\text{PLAY}^{+}$-CS.
In Section \ref{sec:mmv}, we extend the PLAY-CS algorithm to deal with joint signal reconstruction problem in Multiple Measurement Vectors (MMV) setting.
Simulation results for dynamic channel tracking and BSCM are provided in Section \ref{sec:ce} and \ref{sec:bscm}, respectively. 
Finally, the paper concludes in Section \ref{sec:conclusion}.

\textit{Notation}: The notation $\|x\|_k$ denotes the $\ell_k$ norm of the vector $x$. $A^{-1}, A^{\prime}$ and $A^H$ denote the inverse, transpose and conjugate transpose of matrix $A$, respectively. {For a set $T$, we use} $T^c$ to denote the complement of $T$. $|T|$ denotes the cardinality of the set $T$. We use the notation $A_T$ to denote the sub-matrix containing the columns of $A$ with indexes belonging to $T$. For a vector $x$, the notation $(x)_T$ refers to a sub-vector that contains the elements with indexes in $T$.

\section{Problem Formulation} \label{sec:problem}
\subsection{Problem Definition}
The main goal of {DCS} problem is to recursively reconstruct a high dimensional sparse $N$-length vector signal sequences $\{x_t\}$ from potentially noisy and undersampled $M$-length measurements $\{y_t\}$(i.e., $M\ll N$) satisfying
\begin{equation}
	y_t = A_tx_t+n_t,n_t\sim \mathcal{CN}(\boldsymbol{0},R_t),
	\label{dcs}
\end{equation}
where $A_t: = G_t\Phi \in \mathbb{C}^{M\times N}$ and $n_t$ is a complex Gaussian noise vector with covariance $R_t$. Here $G_t$ is the measurement matrix and $\Phi$ is a {dictionary matrix} for the sparsity basis. In the formulation above, $z_t:= \Phi x_t$ is actually the signal of interest at time $t$, where $x_t$ is its representation in the sparsity basis $\Phi$.

\subsection{System Model}
In this paper, we focus on the time-varying signal sequences that have the following dynamic filtering model:
\begin{equation}
	x_t = f_t(x_{t-1}) + \nu_t ,
\end{equation}
where $f_n(\cdot): \mathbb{C}^N \to \mathbb{C}^N$ is the dynamic function and $\nu_t$ is the filtering noise (innovation) that represents the evolving noise in the dynamic model $f_n(\cdot)$.

Most standard dynamic filtering techniques are based on the Kalman filter, which often models the dynamic function as a linear version and assumes the filtering noise $\nu_t$ is a complex Gaussian noise vector with covariance $Q_t$:
\begin{equation}
	x_t = F_t x_{t-1} + \nu_t,\nu_t \sim \mathcal{CN}(\boldsymbol{0},Q_t),
	\label{dm-l-g}
\end{equation}
where $F_t\in \mathbb{C}^{N\times N}$ is the linear version of the dynamic function $f_n(\cdot)$.

In our work, we propose the {Partial-Laplacian sparsity model}, 
which extends model \eqref{dm-l-g} to a more general case. 
Building upon this new model, we propose a novel DCS algorithm and elucidate its connections to many existing algorithms. In other words, our proposed algorithm can boil down to several previously established algorithms under specific conditions, thereby highlighting the contributions of our work to the study of DCS algorithms.

\section{Applications} \label{sec:application}
Our application focuses on two primary tasks for detecting and tracking objects in wireless sensor networks and video sequences: dynamic channel tracking \cite{mimo} and BSCM \cite{BSCM_2016}.

\begin{figure*}[!t]
	\centering
	\includegraphics[width=4in]{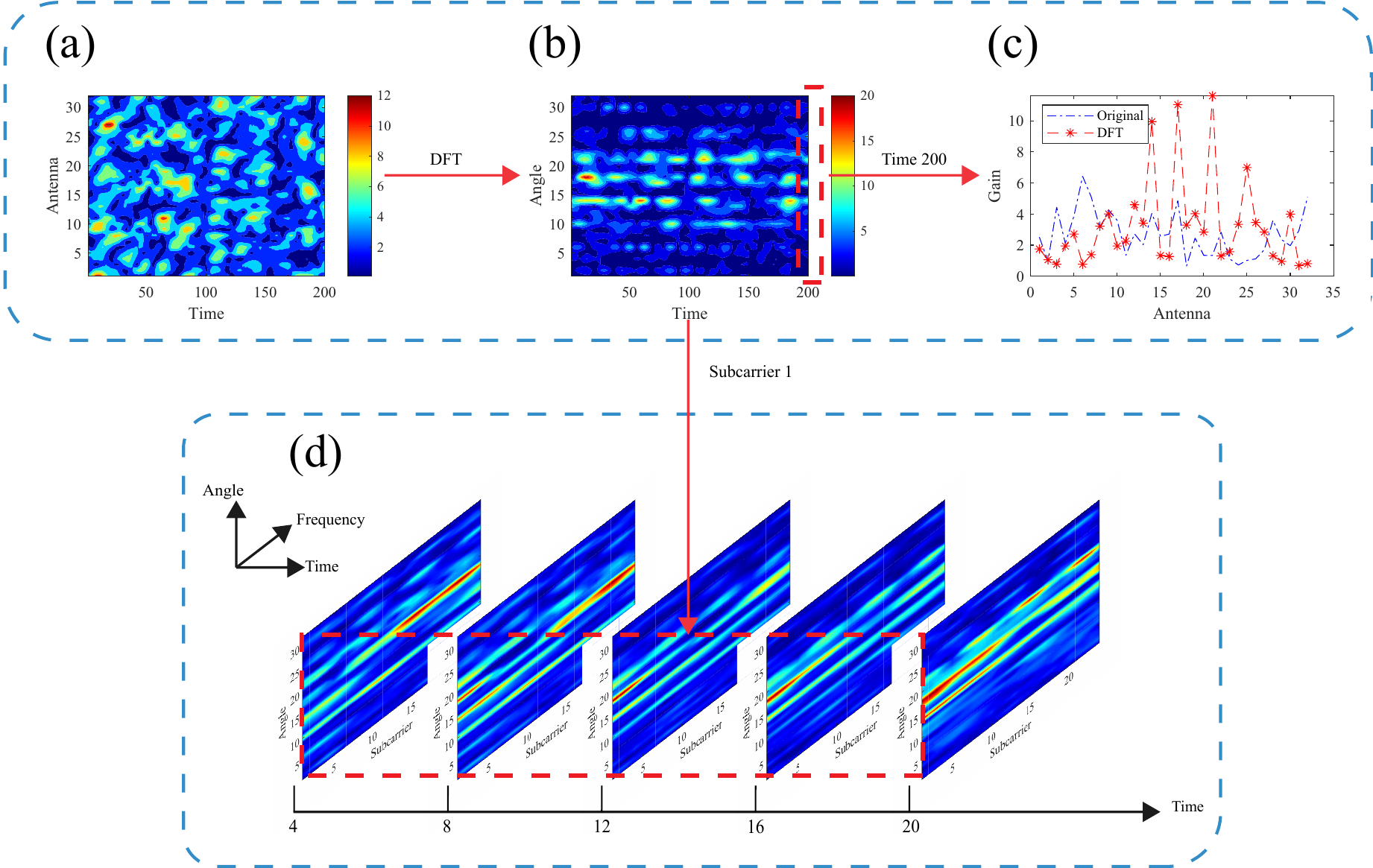}
	\caption{{An example of the CDL-B channel model in \cite{cdl}. (a) plots the channel gains in the antenna-time domain. (b) plots the channel gains in the angle-time domain. (c) plots the sparsity in angular domain. (d) shows the dynamic joint sparsity in the angle-frequency-time domain.}}
	\label{fig_channel_analysis}
\end{figure*}

\subsection{{Dynamic Channel Tracking}}
We {first} consider a {narrowband} massive MIMO system with {HBF} \cite{hbf-lt},
where the base station (BS) is equipped with $N_{\text{r}} \gg 1$ antennas and $N_{\text{RF}}$  {radio frequency (RF)} chains, where $N_{\text{r}} \gg N_{\text{RF}}$.
The channel ${h_t}\in \mathbb{C}^N$ at each time slot $t$ is precoded through an $N_{\text{RF}} \times N_{\text{r}}$ analog beamforming $W_t$, which can be generated by discrete Fourier transform (DFT) matrix.

Due to the {scattering} characteristic of the mmWave propagation \cite{hbf-ssp}, the channel is modeled with $N_L$ paths. Considering the mmWave system equipped with a half-wave spaced uniform linear array (ULA) at the receiver, the channel vector ${h_t}$ can be represented as
\begin{equation}
	{h_t} = \sum_{i=1}^{N_L} \alpha_i \text{a}(\theta_i^t),
\end{equation}
where $\alpha_i$ represents the complex gain of the $i$-th path and $\text{a}(\theta_i^t)$ denotes the normalized responses of the receive antenna arrays to the $i$-th path at time slot $t$:
\begin{equation}
	\text{a}(\theta_i^t) = \frac{1}{\sqrt{N_{\text{r}}}} [1\ \ e^{j \pi \sin{\theta_i^t} }\ \ ...\ \ e^{j \pi (N_{\text{r}}-1)\sin{\theta_i^t} } ]^T,
\end{equation}
where $\theta_i^t$ denotes the angle of arrival.

The observed precoded channel ${y_t} \in \mathbb{C}^{N_{\text{RF}}}$ at time slot $t$ can be represented as 
\begin{equation}
	{y_t} = W_t {h_t} + n_t,n_t\sim \mathcal{CN}(\boldsymbol{0},\sigma_m^2 {I}_{N_{\text{RF}}}),
	\label{eq_channel_mea}
\end{equation}
where $n_t$ is a complex Gaussian noise vector with covariance $\sigma_m^2 {I}_{N_{\text{RF}}}$. Noticing the {scattering} structure of the channel, ${h_t}$ can be transformed into the sparse angle domain, i.e., ${h_t}$ can be expressed as 
\begin{equation}
	{h_t} = \text{D} {\tilde{h}_t},
	\label{eq_channel_sparse}
\end{equation}
{where $\text{D}$ is the transform dictionary determined by the geometrical structure of the antenna array, and ${\tilde{h}_t}$ is the sparse representation of the channel in the angle domain.}
The channel sequences $\{{\tilde{h}_t}\}$ in the angle domain exhibit strong structured dynamic sparsity, as depicted {in Fig. \ref{fig_channel_analysis} (a), (b) and (c)}. {In this paper, we focus on the ULA at {BS},} where D is the {DFT matrix}. Our discussion can be readily extended to a uniform planner array (UPA) or a more sophisticated antenna array. Substituting \eqref{eq_channel_sparse} and letting $A_t = W_t \text{D}$, we can rewrite \eqref{eq_channel_mea} as
\begin{equation}
	{y_t} = A_t {\tilde{h}_t} + n_t,
\end{equation}
which has the same problem formulation as the DCS problem \eqref{dcs}. Our goal is to recursively reconstruct the uplink channel sequences $\{{h_t}\}$ from the low-dimensional observed precoded channel sequences $\{{y_t}\}$.

\begin{figure}[!t]
	\centering
	\includegraphics[width=3in]{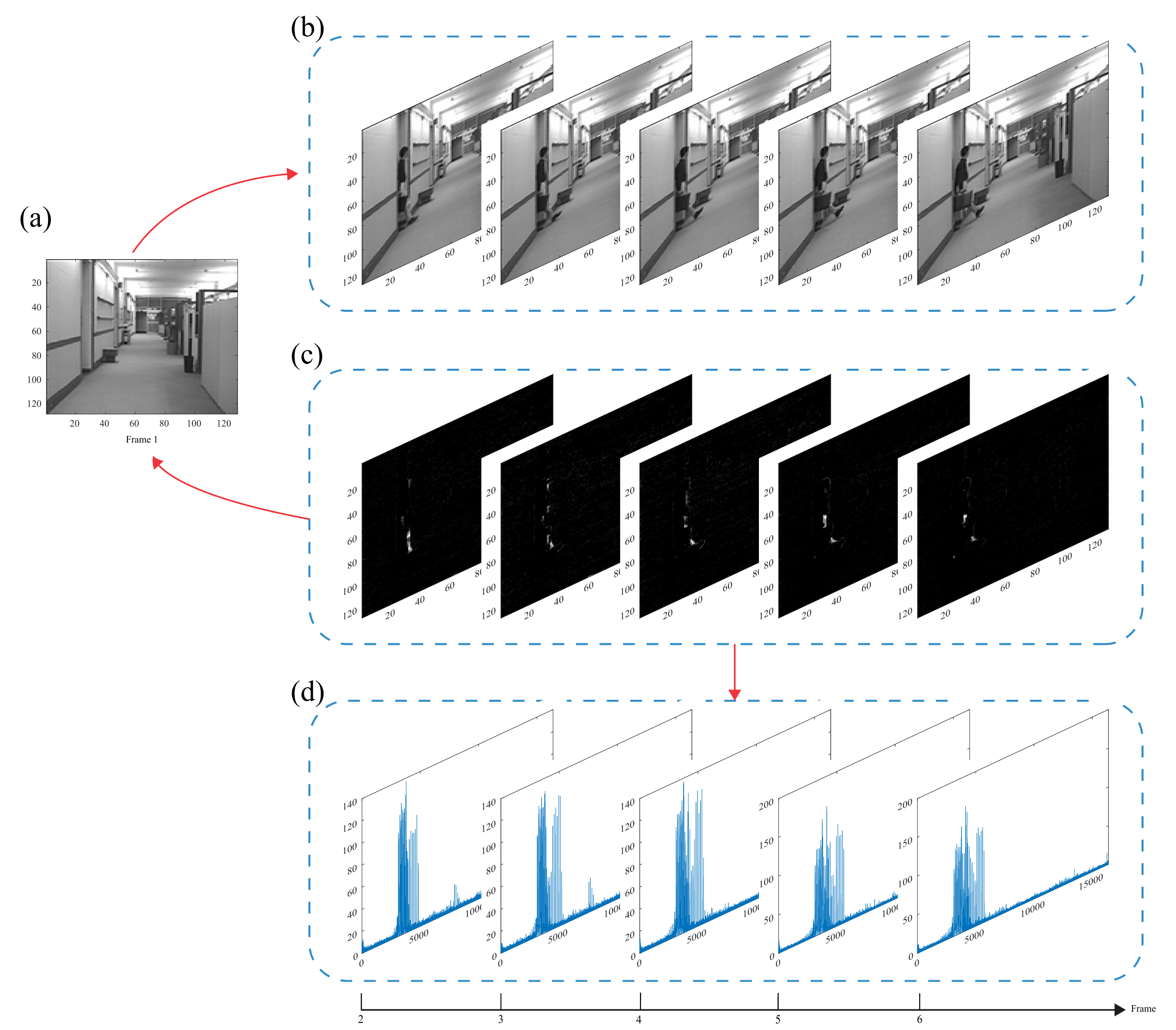}
	\caption{{Consecutive frames in the Hall video sequence \cite{hall}. (a) plots the background (frame 1). (b) plots the video frames. (c) plots the subtracted foregrounds. (d) shows the vectorized pixel values of the corresponding foregrounds.}}
	\label{fig_video_analysis}
\end{figure}

\subsection{Background Subtraction from Compressive Measurements}
Background subtraction, a fundamental research area in computer vision over recent decades \cite{BSCM_2004,BSCM_2010,BSCM_2014}, involves the simultaneous extraction of video backgrounds and segmentation of moving objects within video frames. Conventional approaches typically entail dense scene sampling, necessitating significant storage resources.
Motivated by the advances in CS, a novel technique for background subtraction, termed BSCM \cite{BSCM_2016}, has emerged, promising to alleviate computational burdens and enhance efficiency. 

We now begin to model the BSCM problem as a DCS problem.
Let $\{Z[k]\}_{k\geq1}$ be a sequence of images with resolution $N_1\times N_2$,
and let $z[k] = \operatorname{vec}(Z[k])\in \mathbb{R}^n$ with $n=N_1\cdot N_2$.
At time instance $k$, we observe the compressive measurement $u[k] = A_k z[k] + n[k]$, 
where $A_k$ is a measurement matrix
and $n[k]$ is a Gaussian noise vector.
We decompose each image $Z[k]$ as $Z[k] = X[k] + B$, 
where $X[k]$ is the {$k$-th} foreground image, typically sparse,
and $B$ is the background image, assumed known and the same in all images.
Let $x[k] = \operatorname{vec}(X[k])$ and $b = \operatorname{vec}(B)$.
Because the background image is known, we take measurements from it using $A_k$: $u^b[k] = A_k b$.
Then, we have
\begin{equation}
	y[k] = u[k] - u^b[k] = A_k(z[k]-b) + n[k] = A_k x[k] + n[k].
	\label{eq:prob_bscm}
\end{equation}
Problem \eqref{eq:prob_bscm} aims at recovering the full videos from relatively fewer measurements, constituting a severely ill-posed inverse problem.
Thus deeply exploiting the video structure inforemation is essential to making the problem well-posed.
From {Fig. \ref{fig_video_analysis}}, we can observed that the foreground frames exhibit strong sparsity and temporal correlation, offering avenues for enhancing recovery accuracy.

\section{Algorithmic Framework} \label{sec:algorithm}

\subsection{Kalman Filter}
The Kalman filter \cite{kfcs} reformulates the above problem as causal minimum mean squared error (MMSE) estimation. If the support is {set}, the Kalman filter provides the MMSE solution. {In the Kalman filter framework, the signal at each time step is recovered using the estimate of previous time step $\hat{x}_{t-1}$ and a calculated covariance for that estimate $P_{t-1}$.} {The Kalman filter is typically depicted as comprising two phases: prediction and update.}

{Prediction}:
\begin{align}
	\hat{x}_{t|t-1} &= F_t\hat{x}_{t-1},\\
	P_{t|t-1} &= F_tP_{t-1}F_t^H + Q_t,
\end{align}
where $P_t$ is $\mathbb{E}[ (x_t - \hat{x}_t)(x_t - \hat{x}_t)^H]$ and $P_{t|t-1}$ is $\mathbb{E}[ (x_t - \hat{x}_{t|t-1})(x_t - \hat{x}_{t|t-1})^H]$.

{Update}:
\begin{align}
	K &= P_{t|t-1}A_t^H(A_tP_{t|t-1}A_t^H + R_t)^{-1},\\
	\hat{x}_t &= \hat{x}_{t|t-1} + K(y_t - y_{t|t-1}), y_{t|t-1}=A_t\hat{x}_{t|t-1},\\
	P_t &= (I - KA_t)P_{t|t-1},
\end{align}
where $K$ is the Kalman gain.

When it is assumed that measurement noise $R_t$ and filtering noise $Q_t$ follow a Complex Gaussian distribution, the MMSE estimate of the signal is equivalent to the maximum a posteriori (MAP) estimate. Using the Bayes' rule we have $p(x_t|y_t) \propto  p(y_t|x_t)p(x_t)$, and therefore the MAP estimate $\hat{x}_{t}$ is given by
\begin{equation}
	\hat{x}_{t} = \arg \min_x \{ \|y_t - A_t x\|_{R_t^{-1}}^2 + \| x - \hat{x}_{t|t-1}\|_{P_{t|t-1}^{-1}}^2 \},
	\label{kf-opt}
\end{equation}
where the matrix weighted norm is defined as {$\|z\|_R^2 = z^HRz$}. 
The optimization described in \eqref{kf-opt} effectively utilizes local information, incorporating parameters like the covariance matrix $P_{t|t-1}$ and the previous state estimate $\hat{x}_{t|t-1}$ to enhance the estimation process. However, reliance on the covariance matrix assumes Gaussian and linear characteristics, which may not hold in scenarios where signals and innovations deviate from Gaussian statistics \cite{rwl1-df}.

In cases where the linearity and Gaussianity conditions are not met, {it is necessary to extend the Kalman filter to a more general algorithmic framework}. In this study, we firstly propose the Partial-Laplacian filtering sparsity model. Moreover, we found the new algorithmic framework based on the Partial-Laplacian model has a close relationship with the existing {DCS algorithms \cite{kfcs,modcs,regmodbpdn,rwl1-df}}. {It means that} the new framework can boil down to several existing DCS algorithms under certain conditions.

\begin{algorithm}[t]
	\renewcommand{\algorithmicrequire}{\textbf{Input:}}
	\renewcommand{\algorithmicensure}{\textbf{Output:}}
	\renewcommand{\algorithmicprint}{\textbf{Initialize:}} 
	\caption{{Partial Laplacian Dynamic CS (PLAY-CS)}}\label{alg:alg1-2G-laplacian}
	\begin{algorithmic}
		\REQUIRE { $\{y_1,y_2,...,{y_T}\}$, $A_t,\forall t$, $\sigma_m^2$, {$\sigma_f^2$}, $\alpha$, $F_t,\forall t$, {$W_t,\forall t$}}
		\PRINT $Q_t = \sigma_m^2 {I}, R_t = \sigma_f^2 I, \forall t$, {$\lambda = \frac{\sigma_m^2}{\sigma_f^2}$}, $P_0 = I, \hat{x}_0=\boldsymbol{0},T_0=\emptyset$
		\FORALL {$t=1,2,...,T$}
		\STATE {\textbf{Prediction}}
		\STATE \hspace{0.5cm} $\hat{x}_{t|t-1} = F_t\hat{x}_{t-1},$
		\STATE \hspace{0.5cm} $P_{t|t-1} = F_tP_{t-1}F_t^H + Q_t,$
		\STATE \hspace{0.5cm} $T = T_{t-1}.$
		\STATE {\textbf{Update}}
		\STATE \hspace{0.5cm} $K = P_{t|t-1}A_t^H(A_tP_{t|t-1}A_t^H + R_t)^{-1}.$
		\STATE \hspace{0.5cm} Estimate $\hat{x}_t$ using \eqref{kf-lp-opt}.
		\STATE \hspace{0.5cm} $P_t = (I - KA_t)P_{t|t-1}.$
		\STATE \hspace{0.5cm} Support estimation: $T_t = \{i:|(\hat{x}_t)_i|>\alpha\}$.
		\ENDFOR
		\ENSURE $\{\hat{x}_1,\hat{x}_2,...,{\hat{x}_T}\}$
	\end{algorithmic}
\end{algorithm}

\subsection{Partial-Laplacian Filtering Sparsity Model}
Considering the {slow-changing} characteristic of the signal sequences, which means the values outside the support evolve very sparsely. 
Inspired by the CS technique, one important issue is to capture the sparsity of {the addition to the support} with a proper regularization term, such as $\ell_1$ norm. 

To capture the dynamic sparsity outside the support, we propose the Partial-Laplacian filtering sparsity model:
\begin{equation}
	\begin{aligned}
		(x_t)_{T_{t-1}} &= (F_tx_{t-1})_{T_{t-1}} + (\nu_t)_{T_{t-1}},\\
		(x_t)_{T_{t-1}^c} &= (F_tx_{t-1})_{T_{t-1}^c} + (\nu_t)_{T_{t-1}^c},
	\end{aligned}
	\label{Bipartite-filter-noise}
\end{equation}
{where $T_{t}$ denotes the the support set of $x_{t}$} and let {$L = |T_t|$}. In other words, $T_{t}=[i_1, i_2, ..., i_{L}]$, where $\{i_k\}_{k=1}^{k=L}$ are the non-zero coordinates of $x_t$. {We assume that $(\nu_t)_{T_{t-1}}\sim \mathcal{CN}(\boldsymbol{0},Q_t^1)$}
and $(\nu_t)_{T_{t-1}^c}$ have independent Laplacian but non-identical distributions with inverse scale $w_i$, i.e. $p((\nu_t)_i) = \frac{w_i}{2} e^{-w_i|(\nu_t)_i|},i \in T_{t-1}^c$. 

Based on the dynamic model \eqref{Bipartite-filter-noise}, the MAP estimate is
\begin{equation}
	\begin{aligned}
		\hat{x}_{t} = \arg \min_x&  \{ \|y_t - A_t x\|_{R_t^{-1}}^2 + {\lambda} \| (x)_T - (\hat{x}_{t|t-1})_T\|_{(P_{t|t-1})_1^{-1}}^2 \\
		&+ \|{W}_t((x)_{T^c}  - (\hat{x}_{t|t-1})_{T^c}) \|_1\},
		\label{kf-lp-opt}
	\end{aligned}
\end{equation}
where the submatrix $(P_{t|t-1})_1 = {\lambda} (P_{t|t-1})_{T,T}$ 
\footnote{{Determining the appropriate value of parameter $\lambda$ is inherently challenging, and no universal method exists for its selection. In this paper, we adopt a strategy similar to that in \cite{regmodbpdn} for determining $\lambda$. Specifically, we assume that  $n_t\sim \mathcal{CN}(\boldsymbol{0},\sigma_m^2 {I})$ and $(\nu_t)_{T_{t-1}} \sim \mathcal{CN}(\boldsymbol{0},\sigma_f^2 {I})$. Based on these assumptions, we set $\lambda = \frac{\sigma_m^2}{\sigma_f^2}$.}} 
and $W_t={\text{diag}} (w_{1}, w_{2}, ..., w_{N-L})$ is a diagonal matrix.

When we set {$R_t=2I$, $(P_{t|t-1})_1=2I$}, the optimization problem \eqref{kf-lp-opt} will be as following
\begin{equation}
	\begin{aligned}
		\hat{x}_{t} = \arg \min_x&  \{ \frac{1}{2} \|y_t - A_t x\|_{2}^2 + \frac{1}{2} \lambda \| (x)_T - (\hat{x}_{t|t-1})_T\|_{2}^2 \\
		&+ \|W_t((x)_{T^c}  - (\hat{x}_{t|t-1})_{T^c}) \|_1\}.
		\label{kf-lp-opt-2norm}
	\end{aligned}
\end{equation}

Utilizing the Partial-Laplacian filtering sparsity model as a foundation, we can formulate a comprehensive DCS algorithmic framework {called the Partial Laplacian Dynamic CS (PLAY-CS)}, as outlined in Algorithm \ref{alg:alg1-2G-laplacian}. This newly developed DCS algorithm can transform into pre-existing DCS algorithms \cite{kfcs,modcs,regmodbpdn,rwl1-df}, a demonstration of which will be presented in the subsequent section.

\subsection{Connection to Other DCS Algorithms} \label{subsec_connection}

\begin{table*}[]
	\caption{Summary of the connection between PLAY-CS and existing DCS algorithms \label{tab:comprasion-dcs}}
	\centering
	\begin{tabular}{|c|c|}
		\hline
		Algorithm  & Connection with PLAY-CS                                      \\ \hline
		KF-CS \cite{kfcs}     & $(x_t)_T$ and $(x_t)_{T^c}$ independent, $W_t=I$     \\ \hline
		Modified-CS \cite{modcs}  & $W_t=I,{\hat{x}_{t|t-1}=0},\gamma=0$                               \\ \hline
		Reg-mod-BPDN \cite{regmodbpdn}   & $W_t=I,(\hat{x}_{t|t-1})_{T^c}=0$    \\ \hline
		Weighted-$\ell_1$ \cite{rwl1-df}  & $\hat{x}_{t|t-1}=0$,$T=\emptyset$                                           \\ \hline
	\end{tabular}
\end{table*}

The proposed {PLAY-CS} algorithm can reveal the intrinsical correlations of some existing DCS algorithms \cite{kfcs,modcs,regmodbpdn,rwl1-df,Mota_2016_TSP,Luong_2016_ICIP}. Table \ref{tab:comprasion-dcs} shows the connections between {PLAY-CS} algorithm and some existing DCS algorithms.

\textbf{KF-CS \cite{kfcs}}:
Introduced in \cite{kfcs}, KF-CS algorithm is employed to solve a recursive DCS problem.
The key idea is to run a KF in signals' estimate support, and then execute {an additional detection} outside the support by solving a $\ell_1$ minimization problem.
We mention that KF-CS can be viewed as a special case of the {PLAY-CS} under assumption that the filtering noise $(\nu_t)$ can be divided into two independent parts: $(\nu_t)_{T}$ and $(\nu_t)_{T^c}$ and {$W_t = I$}. This means {the {PLAY-CS}} can be executed through two consecutive steps, which is equivalent to the KF-CS algorithm.

{\textbf{Modified-CS \cite{modcs}}:} For the noisy measurement, the solution of the Modified-CS problem\cite{modcs} is given by
\begin{equation}
	\hat{x}_t = \arg \min_{x} \|{y}_{t} - A_t x \|_2^2  + \|(x)_{T^c}\|_1.
	\label{modcs}
\end{equation}
Assuming $\lambda=0$ and $W_t=I$ in \eqref{kf-lp-opt}, the Modified-CS problem can be viewed as a special case of the {PLAY-CS}.

\textbf{{Reg-mod-BPDN} \cite{regmodbpdn}}: The Algorithm in \cite{regmodbpdn} furtherly exploit the slow value changing structure of the DCS problem by adding a slow value change penalty item to {the Modified-CS}
\begin{equation}
	\hat{x}_t = \arg \min_{x} {\frac{1}{2}}\|{y}_{t} - A_t x \|_2^2 + \frac{1}{2} \lambda \| (x)_T - {(\hat{\mu})_T}\|^2 + \gamma \|(x)_{T^c}\|_1,
	\label{regmodbpdn}
\end{equation}
which can be derived based on the {PLAY-CS} under the assumption that $W_t=\gamma I$ and $(\hat{x}_{t|t-1})_{T^c} = \boldsymbol{0}$. We will show that the problem presented in \eqref{kf-lp-opt-2norm} can also be interpreted as a optimization problem \eqref{regmodbpdn} {in a sense}.
Letting $\beta = x - \hat{x}_{t|t-1}$, the problem \eqref{kf-lp-opt-2norm} is equivalent to the following problem
\begin{equation}
	\begin{aligned}
		&\hat{\beta}_{t} = \arg \min_{\beta}  \{ \frac{1}{2} \|\tilde{y}_t - A_t \beta\|_{2}^2 + \frac{1}{2} {\lambda} \| (\beta)_T \|_{2}^2 + \|W_t(\beta)_{T^c} \|_1\}, \\
		&\hat{x}_{t} = \hat{\beta}_{t} + \hat{x}_{t|t-1},
		\label{kf-lp-opt-2norm-beta}
	\end{aligned}
\end{equation}
where $\tilde{y}_t = y_t - A_t\hat{x}_{t|t-1}$. The task of solving $\hat{\beta}_{t}$ in \eqref{kf-lp-opt-2norm-beta} can be regarded as a specific instance of the problem presented in \eqref{regmodbpdn}.

\textbf{Weighted-$\ell_1$ \cite{rwl1-df}}:
Consider the following Weighted-$\ell_1$ problem
\begin{equation}
	\hat{x}_t = \arg \min_{x} \frac{1}{2} \|{y}_{t} - A_t x \|_2^2  + \|W_tx\|_1,
	\label{weighted-l1}
\end{equation}
where $W_t={\text{diag}} (w_{1}, w_{2}, ..., w_{N})$. Weighted-$\ell_1$ can be viewed as an extension of Modified-CS. Moreover, Weighted-$\ell_1$ can be considered a simplified variant of the {PLAY-CS}. {Notably, it shares a similar interpretation with Modified-CS when we set $T=\emptyset$.}
Furthermore, the subproblem of solving $\hat{\beta}_{t}$ in \eqref{kf-lp-opt-2norm-beta} can be regarded as a specialized Weighted-$\ell_1$ problem:
\begin{equation}
	\hat{\beta}_{t} = \arg \min_{\beta}  \{ \frac{1}{2} \|\bar{y}_t - \bar{A}_t \beta\|_{2}^2 + \|W_t(\beta)_{T^c} \|_1\},
	\label{eq_beta_weight}
\end{equation}
where $\bar{y}_t = \begin{bmatrix}
	\tilde{y}_t \\
	{\boldsymbol{0}}
\end{bmatrix}$, and 
$\bar{A}_t=\begin{bmatrix}
	A_t \\
	{\sqrt{{\lambda}}I_{T}^{{\prime}}}
\end{bmatrix}$.

\subsection{Partial-LSM Filtering Sparsity Model}

\begin{figure}[!t]
	\centering
	\includegraphics[width=2.7in]{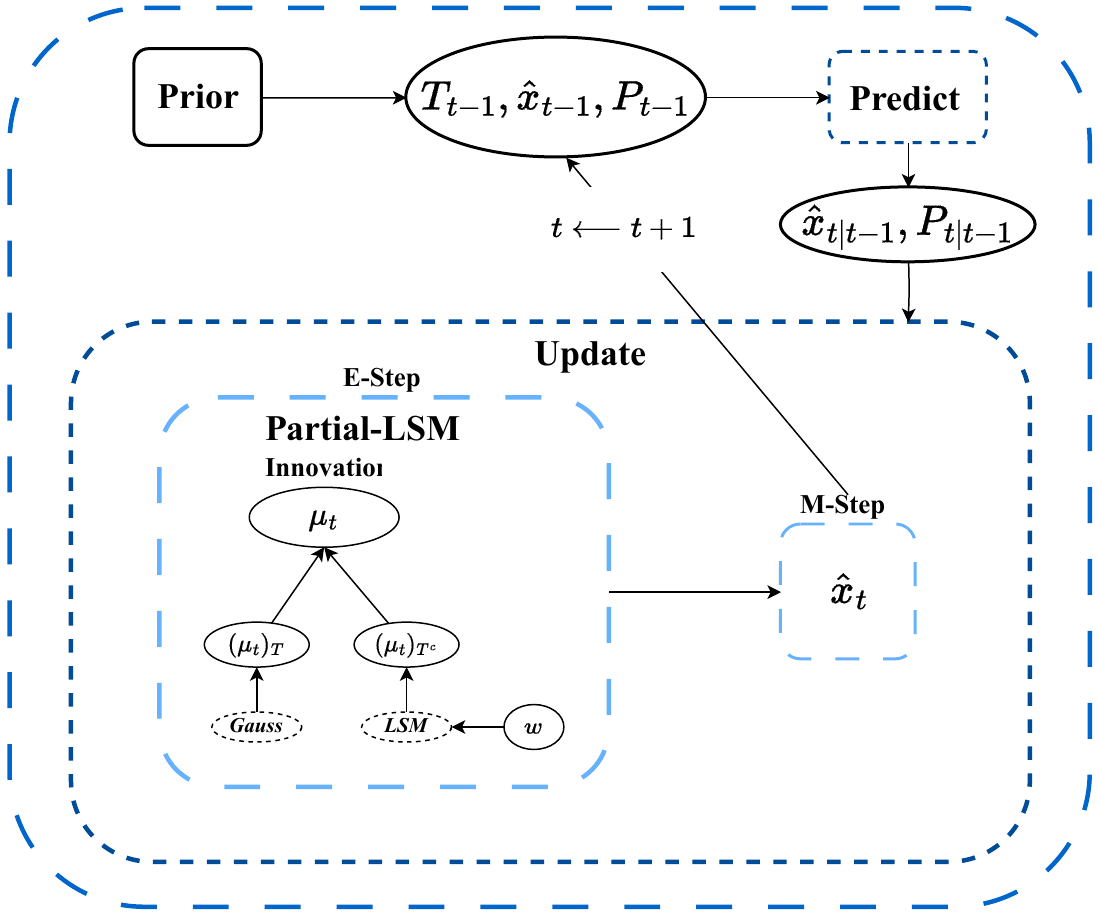}
	\caption{The hierarchical structure of the Partial-LSM model.}
	\label{fig_dcs_flow}
\end{figure}

Although Partial-Laplacian model is useful for illustrating the signal's structured dynamic sparsity, parameter $W_t$ tuning remains a difficult problem for the proposed algorithm. {To address the above issue}, the Partial Laplacian scale mixture (Partial-LSM) filtering sparsity model shown as {Fig. \ref{fig_dcs_flow}} is proposed in this study.

Firstly, we shall introduce the Laplacian scale mixture (LSM) \cite{LSM-nips2010} sparse prior. Given the vector { $\lambda$}, LSM assumes the signal {$x$} have independent but non-identically distributed Laplacian entries, i.e., $p(x\mid \lambda)=\prod_{n=1}^N p\left(x_n \mid \lambda_n\right)$, where
\begin{equation}
	p\left(x_n \mid \lambda_n\right) = \frac{\lambda_n}{2} e^{-\lambda_n|x_n|},\forall n,
\end{equation}
and $\lambda_n,\forall n$ are modeled as independent Gamma distributions, i.e., $p(\lambda)=\prod_{n=1}^{N}p(\lambda_n)$ with
\begin{equation}
	p\left(\lambda_n\right)=\Gamma(a,b)=\frac{b^a}{\Gamma(a)} \lambda_n^{a-1} e^{-b \lambda_n},
\end{equation}
where $a,b$ are the shape parameter and inverse scale parameter respectively, which are set to promote sparsity of $x$. With this particular choice of a prior on $\lambda$, the distribution over $x_n$ is computed by integrating out $\lambda_n$ analytically:
\begin{equation}
	p\left(x_n\right)=\frac{a b^a}{2\left(b+\left|x_n\right|\right)^{a+1}},
\end{equation}
which has heavier tails than the Laplacian distribution. Moreover, due to the conjugacy between Gamma distribution and Laplacian distribution, i.e., 
\begin{equation}
	p\left(\lambda_n \mid x_n\right) = \Gamma(\lambda_n;a+1,b+|x_n|),
	\label{lambda-given-x}
\end{equation} 
we can derive an alternatively iterative algorithmic procedure using the EM algorithm \cite{LSM-nips2010}.

However, LSM only models the spatial sparsity of the signal itself. In other words, LSM does not consider the dynamic sparsity of filtering noise outside the support, which motivates us to construct the {Partial-LSM sparsity model} in this study.

In the following, we shall introduce the Partial-LSM sparse prior model to capture the structured dynamic sparsity by modeling the filtering noise outside the support with LSM prior.

In model \eqref{Bipartite-filter-noise}, we further model the inverse scale parameter $w_i$ of $(\nu_t)_i,i \in T_{t-1}^c$ as independent Gamma distributions
\begin{equation}
	p\left( w_i \right)=\frac{b^a}{\Gamma(a)} w_i^{a-1} e^{-b w_i},i \in T_{t-1}^c.
	\label{lsm-w}
\end{equation}

Using Bayes's rule and the assumption that $(x_t)_T$ and $(x_t)_{T^c}$ are independent, the MAP estimate based on the Partial-LSM model is given by
\begin{equation}
	\begin{aligned}
		\hat{x}_t=&\arg \min_x \{ -\log p(x\mid y_t) \}\\
		=& \arg \min_x \{ -\log p(y_t\mid x) - \log p(x) \} \\ 
		=&\arg \min_x \{ -\log p(y_t\mid x) - \log p((x)_T) - \log p((x)_{T^c}) \}
	\end{aligned}\\
\end{equation}

In general, it is difficult to compute the MAP estimate with the Partial-LSM prior on the {filtering noise}. 
However, if we also {know} the latent variable $w=[w_{i_1}, w_{i_2}, ..., w_{i_{N-L}}]^T$, 
we obtain an objective function that can be minimized with respect to the variable $x$. 
The typical approach to addressing such a problem is the EM algorithm.

In the subsequent sections, we will develop an EM algorithm to estimate the MAP value of $x_t$. By applying Jensen’s inequality, we can derive the following upper bound for the posterior likelihood.
\begin{equation}
	\begin{aligned}
		-\log p(x \mid y_t) \leq&-\log p(y_t \mid x)- \log p((x)_T) \\
		& - \int_w q(w) \log \frac{p((x)_{T^c}, w)}{q(w)} d w:=\mathcal{L}(q, x),
	\end{aligned}
	\label{jensen-ineq}
\end{equation}
which is true for any probability distribution $q(w)$. Employing the EM algorithm as the foundation, we can execute coordinate descent in $\mathcal{L}(q, x)$. This process involves two pivotal updates, commonly referred to as the E step and the M step:
\begin{align}
	\text { E Step } \quad & q^{(k+1)}=\underset{q}{\arg \min } \mathcal{L}\left(q, x^{(k)}\right) \label{E-step},\\
	\text { M Step } \quad & x^{(k+1)}=\underset{x}{\arg \min } \mathcal{L}\left(q^{(k+1)}, x\right), \label{M-step}
\end{align}
where $k$ denotes the $k$-th iteration step in estimation of $x_t$ 

Let $\langle.\rangle_q$ denote the expectation with respect to $q(w)$. The M Step \eqref{M-step} simplifies to \eqref{kf-lp-opt}, where $(W_t)^k={\text{diag}}(\langle w_{i_1}\rangle_{q^{k}}, \langle w_{i_2}\rangle_{q^{k}}, ..., \langle w_{i_{N-L}}\rangle_{q^{k}})$. In other words, the weight matrix $(W_t)^k$ can be automatically learned using EM algorithm.

We have equality in the Jensen inequality if $q(w) = p(w\mid x)$. The inequality \eqref{jensen-ineq} is therefore tight for this particular choice of $q$, which implies that the E step reduces to $q^{(k+1)}(w)=p(w\mid x^k)$. Note that in the M step we only need to compute the expectation of $w_i$ with respect to the distribution $q$ in the E step. Hence we only need to compute the sufficient statistics $\langle w_i \rangle_{p(w\mid x^k)}$.

Based on the Partial-LSM model \eqref{lsm-w}, we can use the fact \eqref{lambda-given-x} to compute the sufficient statistics:
\begin{equation}
	\langle w_i \rangle_{p(w\mid x^k)}=\frac{a+1}{b+|(x^k)_i|}
	\label{eq_wii}
\end{equation}

The complete optimization procedure for the Partial-LSM DCS ({$\text{PLAY}^{+}$-CS}) is outlined in Algorithm \ref{alg:alg-2GHS-LSM}. 
We observe that certain parameters in Algorithm \ref{alg:alg-2GHS-LSM}, such as $\alpha$, $a$, $b$ and $F_t$, are crucial for the optimal performance of our proposed algorithms. 
In our implementation, we replace the selection of the threshold $\alpha$ with the choice of the $95$\%-support, as described in \cite{modcs}.
The $\beta$\%-support is defined in Definition \ref{def_percentage} below.
\begin{defn}[\hspace{1sp} \cite{modcs}] \label{def_percentage}
	For compressible signals, we let $N$ be the $\beta$\%-energy support of $x$, i.e., $N := \{i\in[1,N]: x_i^2>\alpha\}$ , where $\alpha $ is the largest threshold for which $N$ contains at least $\beta$\% of the signal energy.
\end{defn}
To make the priors $a$, $b$ non-informative according to \cite{tipping_sbl}, we set their values to small values, specifically $a$=$b$=$10^{-3}$.
The design of the matrix $F_t$ plays a crucial role in the performance of our proposed algorithm.
However, a detailed exploration of the $F_t$ design is beyond the scope of this paper. 
To ensure a fair and direct comparison with existing DCS algorithms, such as those in \cite{rwl1-df}, \cite{modcs}, \cite{regmodbpdn}, we have adopted the identity matrix used in these algorithms.

\begin{algorithm}[t]
	\renewcommand{\algorithmicrequire}{\textbf{Input:}}
	\renewcommand{\algorithmicensure}{\textbf{Output:}}
	\renewcommand{\algorithmicprint}{\textbf{Initialize:}} 
	\caption{{ {PLAY-CS with LSM} ({$\text{PLAY}^{+}$-CS})}}\label{alg:alg-2GHS-LSM}
	\begin{algorithmic}
		\REQUIRE $\{y_1,y_2,...,y_T\}$, $A_t,\forall t$, $\sigma_m^2$, {$\sigma_f^2$}, $\alpha$, $a,b$, $F_t,\forall t$
		\PRINT $Q_t = \sigma_m^2 {I}, R_t = \sigma_f^2 I, \forall t$, {$\lambda = \frac{\sigma_m^2}{\sigma_f^2}$}, $P_0 = I, \hat{x}_0=\boldsymbol{0},T_0=\emptyset$
		\FORALL {$t=1,2,...,T$}
		\STATE {\textbf{Prediction}}
		\STATE \hspace{0.5cm} $\hat{x}_{t|t-1} = F_t\hat{x}_{t-1},$
		\STATE \hspace{0.5cm} $P_{t|t-1} = F_tP_{t-1}F_t^H + Q_t,$
		\STATE \hspace{0.5cm} $T = T_{t-1}.$
		\STATE {\textbf{Update}}
		\STATE \hspace{0.5cm} $K = P_{t|t-1}A_t^H(A_tP_{t|t-1}A_t^H + R_t)^{-1}.$
		\STATE \hspace{0.5cm} {\textbf{E-Step}}
		\STATE \hspace{1cm} {Set diagonal matrix $W_t$ using \eqref{eq_wii}}
		\STATE \hspace{0.5cm} {\textbf{M-Step}}
		\STATE \hspace{1cm} Estimate $\hat{x}_t$ using \eqref{kf-lp-opt}.
		\STATE \hspace{0.5cm} $P_t = (I - KA_t)P_{t|t-1}.$
		\STATE \hspace{0.5cm} Support estimation: $T_t = \{i:|(\hat{x}_t)_i|>\alpha\}$.
		\ENDFOR
		\ENSURE $\{\hat{x}_1,\hat{x}_2,...,\hat{x}_T\}$
	\end{algorithmic}
\end{algorithm}

\section{Extension}\label{sec:mmv}

In this section we point out an immediate extension of PLAY-CS algorithm. Specifically, the proposed algorithm can effectively reconstruct channels in broadband scenarios, leveraging the advantages of MMV.

\subsection{Joint Sparsity and MMV}

Consider the following dynamic joint signal reconstruction problem
\begin{equation}
	Y_t = A_t X_t + Z_t,
	\label{mmv_problem}
\end{equation}
where $Y_t = [y_t^{(1)},\ldots, y_t^{(P)}]$ is the joint measurement matrix, $X_t = [x_t^{(1)},\ldots, x_t^{(P)}]$ represents the joint sparse estimated signal matrix, indicating that the estimated signals $\{x_t^{(p)}\}$ share a common support. Additionally, $Z_t^f = [n_t^{(1)},\ldots, n_t^{(P)}]$ denotes the complex Gaussian noise matrix.

We can independently execute PLAY-CS algorithm $P$ times to reconstruct the signal sequences $\{x_t^{(p)}\}$ from its respective noisy measurement. This scenario is identified as the {Single Measurement Vector (SMV)} problem. However, in the presence of joint sparsity, employing a joint CS estimation approach proves advantageous for enhancing recovery performance at each time slot. This is known as the {MMV problem} and is quite well studied in the literature \cite{mmv1-l1-svd}. A well-known algorithm for MMV problem is $\ell_{2,1}$-norm regularized least squares ($\ell_{2,1}$-LS) \cite{mmv1-l1-svd}
\begin{equation}
	\hat{X}_t = \arg \min_{X} \|{Y}_{t} - A_t X \|_2^2  + {\rho} \|X\|_{2,1},
	\label{mmv_l21}
\end{equation}
{where $\|X\|_2$ is the Frobenius-norm of matrix $X$},
$\|X\|_{2,1} = \sum_{i=1}^{N}\|{X_{i,:}}\|_2$ is the sum of $\ell_2$-norms of rows of $X$ {with ${X_{i,:}}$ denoting the $i$-th row of $X$}, and $\rho>0$ is a regularization parameter. 
The $\ell_{2,1}$-norm regularization is known for promoting row sparsity in the signal matrix $X$. This regularization method imposes a joint sparsity pattern on the supports of the estimated signals, making it a desirable choice in MMV problems where signal samples exhibit joint sparsity.

However, this method fails to leverage the dynamic joint sparse structure inherent in the signal sequences $\{X_t\}$, {as depicted in Fig. \ref{fig_channel_analysis} (d)} . In order to fully exploit the dynamic joint sparse structure of the signal sequences, we propose a DCS adaptation of the $\ell_{2,1}$-LS for the MMV problem. This novel algorithm is denoted as PLAY-CS-MMV.

\subsection{PLAY-CS-MMV}

Considering the applicability of the MMV problem \eqref{mmv_l21} to the joint sparse signals reconstruction problem \eqref{mmv_problem}, a natural idea is to integrate \eqref{mmv_l21} with $\text{PLAY}^{+}$-CS in Algorithm \ref{alg:alg-2GHS-LSM}. This corresponds to the optimization problem presented below
\begin{equation}
	\begin{aligned}
		\hat{X}_{t} = \arg \min_x&  \{ \frac{1}{2} \|Y_t - A_t X\|_{2}^2 + \frac{1}{2} \lambda \| (X)_T - (\hat{X}_{t|t-1})_T\|_{2}^2 \\
		&+ \|W_t((X)_{T^c}  - (\hat{X}_{t|t-1})_{T^c}) \|_{2,1}\},
		\label{play-cs-mmv}
	\end{aligned}
\end{equation}
where $\hat{X}_{t|t-1} = [\hat{x}_{t|t-1}^{(1)},\ldots,\hat{x}_{t|t-1}^{(P)}]$ with $\hat{x}_{t|t-1}^{(p)}$ denoting  the prior value of the source $x_t^{(p)}$. An important point to note is that $T$ and $W_t$ are common support set estimate and weight matrix applicable to all sources $\{x_t^{(p)}\}$.

This raises the immediate question: what values for $T$ and $W_t$ will improve signal reconstruction performance? For example, $T$ and $W_t$ can be set based on the first subcarrier source $x_t^{(1)}$, i.e., that
\begin{equation}
	T_t = \{i:|(\hat{x}_t^{(1)})_i|>\alpha\},
	\label{eq_T_mmv}
\end{equation}
\begin{equation}
	(W_t)_{i,i} =\frac{a+1}{b+|(\hat{x}_{t-1}^{(1)})_i|}.
	\label{eq_w_mmv}
\end{equation}

We will illustrate through experiments that the reconstruction performance of PLAY-CS-MMV is not significantly affected by the choice of the source location $p$.
The PLAY-CS-MMV algorithm with $T$ and $W_t$ set by \eqref{eq_T_mmv} and \eqref{eq_w_mmv} is referred to as $\text{PLAY}^{+}$-CS-MMV.
In the following sections, we will present a comparative analysis of simulation results for our proposed algorithms and other DCS algorithms in practical applications.

\begin{figure*}[t]
	\centering
	\subfloat[]{\includegraphics[width=1.8in]{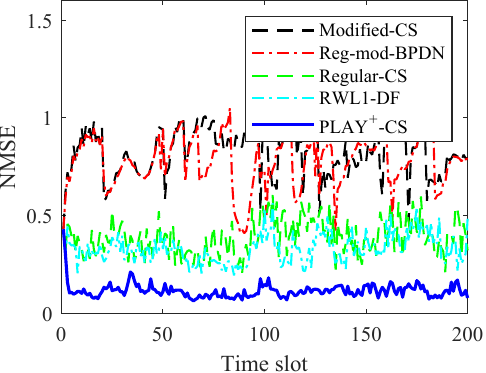}%
		\label{fig_per_time_snr40db_m24_nmse}}
	\hfil
	\subfloat[]{\includegraphics[width=1.8in]{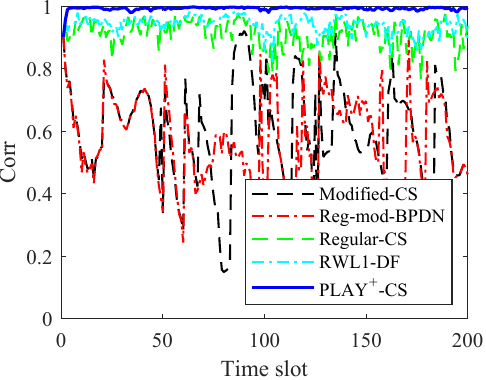}%
		\label{fig_per_time_snr40db_m24_corr}}
	\caption{The NMSE curves and Corr curves of different methods  when SNR of measurements is 40dB and $m=24$. (a) {NMSE} curves. (b) {Corr} curves.}
	\label{fig_per_time_snr40db_m24}
\end{figure*}

\section{{Simulation for Dynamic Channel Tracking}} \label{sec:ce}
\subsection{Experimental Setup}
\textit{Datasets}: In this study, we evaluate the performance of the proposed algorithm using the widely used clustered delay line (CDL) channel models as specified in Table 7.7.1-2 (CDL-B) in \cite{cdl}. A Gaussian random matrix with normalized columns is adopted as the measurement matrix for all methods.

\textit{Comparison Methods}: We compare the proposed method\footnote{{Available at https://github.com/xzliu-opt/PLAY-CS}} with the following state-of-the-art methods: \textbf{Regular-CS} \cite{bp}, \textbf{Modified-CS} \cite{modcs}\footnote{{Available at https://www.ece.iastate.edu/\textasciitilde namrata/modcslargecode.zip}}, \textbf{{Reg-mod-BPDN}} \cite{regmodbpdn}\footnote{{Available at https://www.ece.iastate.edu/\textasciitilde namrata/modcsresidual.zip}}, \textbf{Weighted-$\ell_1$} \cite{rwl1-df}\footnote{{Available at https://www.bme.jhu.edu/ascharles/wp-content/uploads/2020/\\01/RWL1DF\_code.zip}} and \textbf{{$\ell_{2,1}$-LS}} \cite{mmv1-l1-svd}. {Throughout the simulation section, to solve the lasso subproblem of each method, we use the existing solver in the SPGL1 software package \cite{spgl} \footnote{Available at https://friedlander.io/spgl1} or make straightforward modifications therein.}

\textit{Evaluation Measures}: To evaluate the reconstruction performance of the proposed method, {four different metrics} are employed in this study.

\textbf{NMSE}: The resulting normalized mean squared error (NMSE) at each $t$ { was computed by averaging over 100 Monte Carlo simulations of the above model:}
\begin{equation}
	\text{NMSE} := \frac{\|\hat{x}_t - x_t\|^2}{\|x_t\|^2},
\end{equation}

\textbf{Corr}: Another important metric to evaluate the reconstruction performance is the correlation (Corr) between $\hat{x}_t$ and $x_t$ is defined as
\begin{equation}
	\text{Corr} := \frac{\hat{x}_t^H x_t}{\|x_t\| \|\hat{x}_t\|},
\end{equation}
which is a useful performance metric in practical applications, such as channel reconstruction.

\textbf{TNMSE/TCorr}: The average performance metrics that we utilized to analyze the reconstruction performance, which we refer to as the time-averaged NMSE (TNMSE) and time-averaged Corr (TCorr), are defined as
\begin{equation}
	\text{TNMSE} := \frac{1}{T} \sum_{t=1}^{T} \frac{\|\hat{x}_t - x_t\|^2}{\|x_t\|^2},
\end{equation}
\begin{equation}
	\text{TCorr} := \frac{1}{T} \sum_{t=1}^{T} \frac{\hat{x}_t^H x_t}{\|x_t\| \|\hat{x}_t\|},
\end{equation}
where $T$ is the length of the overall signal sequences. Generally, larger Corr, TCorr and smaller NMSE, TNMSE indicate higher reconstruction accuracy.

\begin{figure*}[]
	\centering
	\subfloat[]{\includegraphics[width=1.5in]{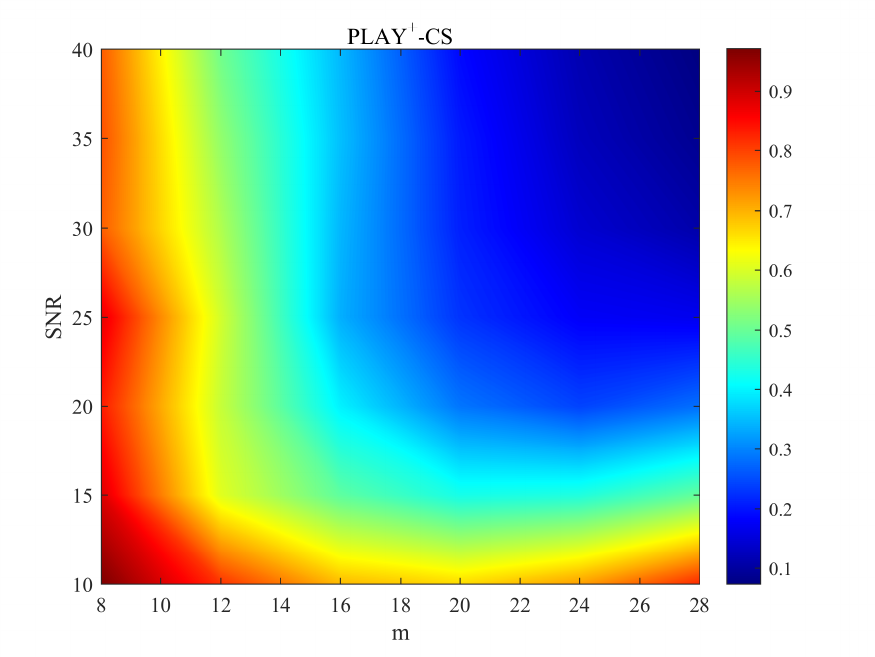}%
		\label{fig_channel_cdl_b_newalg_tnmse}}
	\hfil
	\subfloat[]{\includegraphics[width=1.5in]{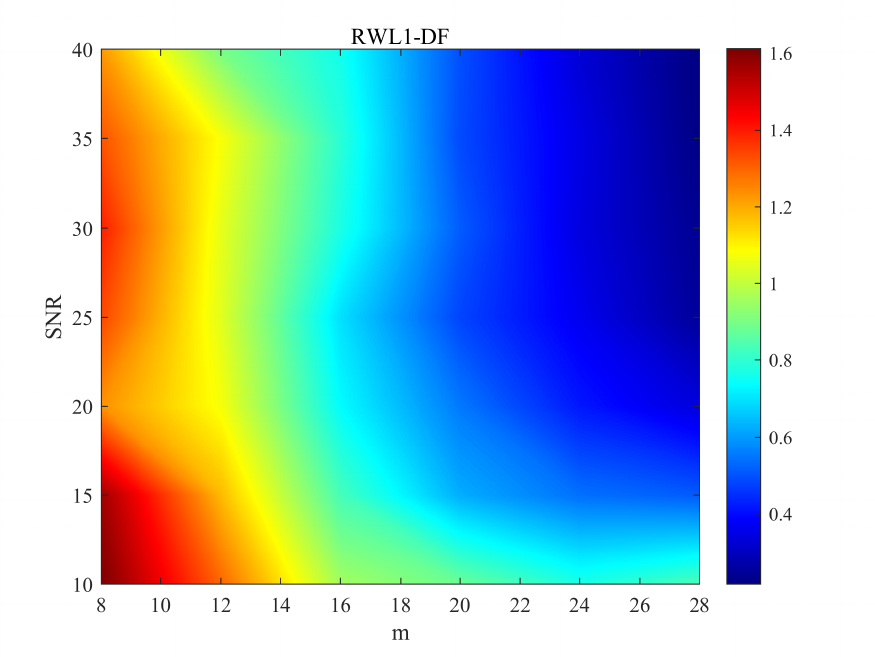}%
		\label{fig_channel_cdl_b_rwl1_df_tnmse}}
	\hfil
	\subfloat[]{\includegraphics[width=1.5in]{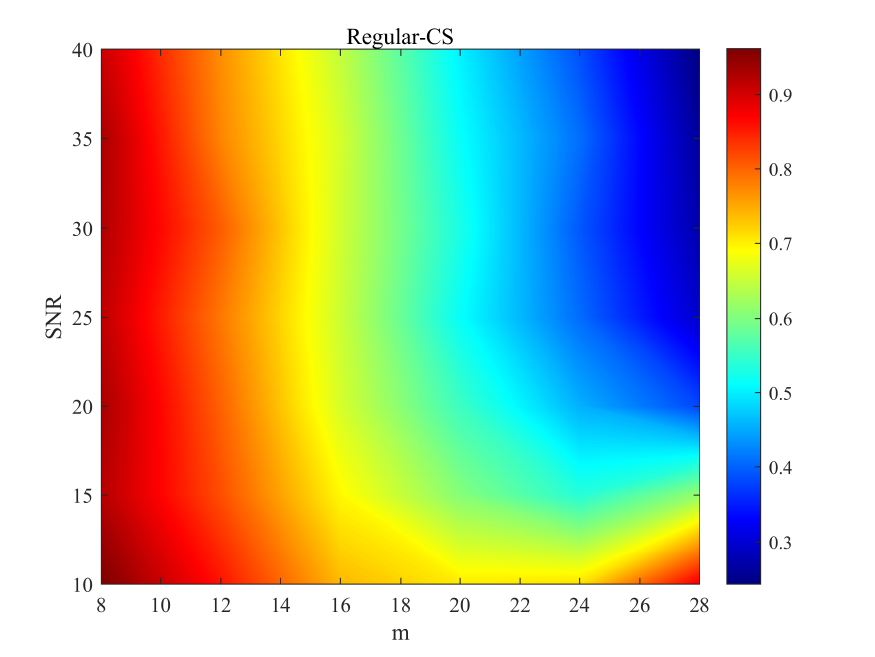}%
		\label{fig_channel_cdl_b_cs_tnmse}}
	\caption{The TNMSE performances of various algorithms under diffirent SNR and CR levels. (a) TNMSE of {$\text{PLAY}^{+}$-CS}. (b) TNMSE of RWL1-DF. (c) TNMSE of Regular-CS.}
	\label{fig_heatmap_tnmse_snr_m}
\end{figure*}

\begin{figure*}[t]
	\centering
	\subfloat[]{\includegraphics[width=1.8in]{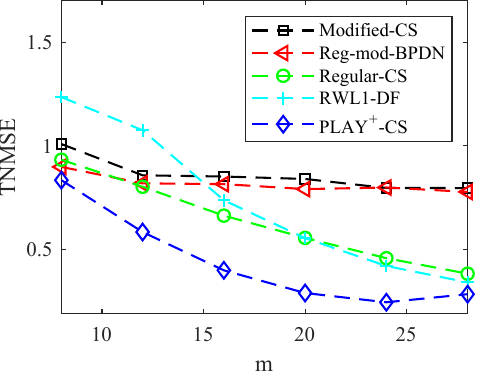}%
		\label{fig_tnmse_m_snr20db}}
	\hfil
	\subfloat[]{\includegraphics[width=1.8in]{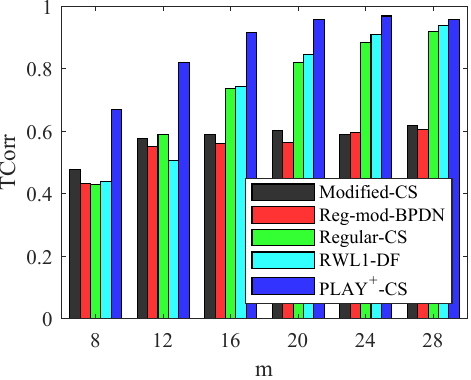}%
		\label{fig_tcorr_m_snr20db}}
	\caption{The TNMSE curves and TCorr bar charts of different methods  when SNR of measurements is 20dB. (a) TNMSE curves. (b) TCorr bar charts.}
	\label{fig_tnmse_tcorr_m_snr20db}
\end{figure*}

\subsection{Performance Comparison in Narrowband Scenarios}

For the channel reconstruction in {narrowband} scenarios, We simulated a 200 consecutive time sequence of {CDL-B dataset}. Under various number of measurement numbers, different levels of additive Gaussian white noise are added into the measurements, which result in the different levels of {compression rate (CR)} (i.e., {the ratio of $m$ to $n$}) and signal noise ratio (SNR) of measurements. {Regular-CS, Modified-CS, {Reg-mod-BPDN}, and RWL1-DF} are implemented as the comparison methods. Given the corresponding noisy measurements $y_t$ and random measurement matrix $A_t$, we recover the signal $\hat{x}_t$ under different levels of CR and SNR.

Figure \ref{fig_per_time_snr40db_m24} illustrates the NMSE and Corr curves versus time slot of various methods when SNR = 40 dB. It is evident that {$\text{PLAY}^{+}$-CS} outperforms other competing DCS methods on TNMSE by at least 4.6 dB on the CDL-B dataset. This improvement is attributed to {$\text{PLAY}^{+}$-CS}'s ability to effectively capture the underlying structured dynamic sparsity in the channel signal sequences.

Compared with RWL1-DF and Regular-CS under a wide range of SNR and number of measurements $m$, the TNMSE performances are shown in Fig. \ref{fig_heatmap_tnmse_snr_m}. From Fig. \ref{fig_heatmap_tnmse_snr_m}, we
see that the proposed {$\text{PLAY}^{+}$-CS} algorithm can achieve a considerable gain over the existing DCS algorithms under various system settings.

\subsection{Impact of SNR}
Table \ref{tab:tnmse_tcorr-snr_m=24} presents the comparison of TNMSE and TCorr scores when $m=24$. The results indicate that {$\text{PLAY}^{+}$-CS} outperforms other methods on the CDL-B dataset, achieving lower TNMSE and higher TCorr scores.

\begin{table*}[]
	\caption{The TNMSE and TCorr scores of different methods when $m=24$  \label{tab:tnmse_tcorr-snr_m=24}}
	\centering
	\begin{tabular}{|ccccc|}
		\hline
		\multicolumn{5}{|c|}{The TNMSE scores}                                                                                                                                                                                              \\ \hline
		\multicolumn{1}{|c|}{}                                               & \multicolumn{1}{c|}{SNR=15} &  \multicolumn{1}{c|}{SNR=25}  & \multicolumn{1}{c|}{SNR=35} & SNR=40 \\ \hline
		\multicolumn{1}{|c|}{Modified-CS\cite{modcs}}       & \multicolumn{1}{c|}{0.5379}  & \multicolumn{1}{c|}{0.4100}  & \multicolumn{1}{c|}{0.4100} & 0.3877 \\
		\multicolumn{1}{|c|}{Reg-mod-BPDN\cite{regmodbpdn}} & \multicolumn{1}{c|}{0.8152}  & \multicolumn{1}{c|}{0.8032} & \multicolumn{1}{c|}{0.7795} & 0.8255 \\
		\multicolumn{1}{|c|}{Regular-CS\cite{bp}}           & \multicolumn{1}{c|}{0.7977}  & \multicolumn{1}{c|}{0.7724}  & \multicolumn{1}{c|}{0.7870} & 0.7759 \\
		\multicolumn{1}{|c|}{RWL1-DF\cite{rwl1-df}}         & \multicolumn{1}{c|}{0.5414}  & \multicolumn{1}{c|}{0.3678}  & \multicolumn{1}{c|}{0.3536} & 0.3307 \\
		\multicolumn{1}{|c|}{\textbf{$\text{PLAY}^{+}$-CS}}                                       & \multicolumn{1}{c|}{\textbf{0.4366}} & \multicolumn{1}{c|}{\textbf{0.1790}}  & \multicolumn{1}{c|}{\textbf{0.1259}} & \textbf{0.1150} \\ \hline
		\multicolumn{5}{|c|}{The TCorr scores}                                                                                                                                                                                              \\ \hline
		\multicolumn{1}{|c|}{}                                               & \multicolumn{1}{c|}{SNR=15} & \multicolumn{1}{c|}{SNR=25}  & \multicolumn{1}{c|}{SNR=35} & SNR=40 \\ \hline
		\multicolumn{1}{|c|}{Modified-CS\cite{modcs}}       & \multicolumn{1}{c|}{0.8426}  & \multicolumn{1}{c|}{0.9019}  & \multicolumn{1}{c|}{0.9080} & 0.9124 \\
		\multicolumn{1}{|c|}{Reg-mod-BPDN\cite{regmodbpdn}} & \multicolumn{1}{c|}{0.5816} & \multicolumn{1}{c|}{0.6048} & \multicolumn{1}{c|}{0.6046} & 0.5953 \\
		\multicolumn{1}{|c|}{Regular-CS\cite{bp}}           & \multicolumn{1}{c|}{0.5831}  & \multicolumn{1}{c|}{0.6053}  & \multicolumn{1}{c|}{0.6244} & 0.5621 \\
		\multicolumn{1}{|c|}{RWL1-DF\cite{rwl1-df}}         & \multicolumn{1}{c|}{0.8595}  & \multicolumn{1}{c|}{0.9253}  & \multicolumn{1}{c|}{0.9344} & 0.9419 \\
		\multicolumn{1}{|c|}{\textbf{$\text{PLAY}^{+}$-CS}}                                       & \multicolumn{1}{c|}{\textbf{0.9098}} & \multicolumn{1}{c|}{\textbf{0.9823}}  & \multicolumn{1}{c|}{\textbf{0.9914}} & \textbf{0.9925} \\ \hline
	\end{tabular}
\end{table*}

\subsection{Impact of CR}

In Fig. \ref{fig_tnmse_tcorr_m_snr20db}, we present a comparative analysis of TNMSE and TCorr performance for various algorithms as a function of the number of measurements, $m$, under noise level with SNR = 20 dB. The proposed {$\text{PLAY}^{+}$-CS} algorithm exhibits a significant performance improvement compared to other DCS algorithms across various values of $m$. This highlights the algorithm's superior capability in effectively tracking dynamic channels in massive MIMO systems by leveraging the channel's structured dynamic sparsity.

\subsection{Benifits of MMV}
\begin{figure}[]
	\centering
	\includegraphics[width=2.8in]{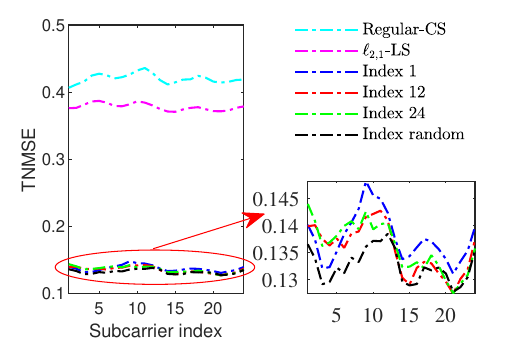}
	\caption{The TNMSE for each subcarrier under various methods when SNR is 20dB, $m=24$ and $P=24$.}
	\label{fig_dcs_mmv}
\end{figure}

In broadband scenarios, to show the benifit of MMV in realistic dynamic channel models of Massive MIMO-OFDM systems \cite{cdl}, we compare the performance of the proposed $\text{PLAY}^{+}$-CS-MMV against Regular-CS \cite{bp} and $\ell_{2,1}$-LS \cite{mmv1-l1-svd}. 
Fig. \ref{fig_dcs_mmv} compares the TNMSE performance of $\text{PLAY}^{+}$-CS-MMV to the baseline algorithms cross the 24 subcarriers.
{Our experiments involved four configurations for selecting the source index, where the first three sets had fixed subcarrier index of 1, 12, and 24, respectively, and the fourth set randomly selected an index at each time slot.} 
The results reveal a significant improvement over the baseline algorithms with different choice of the source index. 
Moreover, this improvement is also fairly robust with respect to the choice of the source index.
This observation implies that $\text{PLAY}^{+}$-CS-MMV can effectively exploit dynamic joint sparsity in broadband scenarios.

\begin{table}[t]
	\centering
	\caption{{Quantitative Results of TNMSE/TCorr with different {weighting scheme}} \label{tab:tnmse_tcorr_ablation_EM}}
	\begin{tabular}{cccc}
		\toprule
		SNR(dB)   & Measures & PLAY-CS & $\text{PLAY}^{+}$-CS \\
		\midrule
		\multicolumn{4}{c}{$m$=16} \\
		\midrule
		\multirow{2}[2]{*}{25} & TNMSE & 0.4731  & \textbf{0.3385} \\
		& Tcorr & 0.8873  & \textbf{0.9381} \\
		\midrule
		\multirow{2}[2]{*}{30} & TNMSE & 0.5025  & \textbf{0.3352} \\
		& Tcorr & 0.8719  & \textbf{0.9417} \\
		\midrule
		\multirow{2}[2]{*}{40} & TNMSE & 0.5302  & \textbf{0.3385} \\
		& Tcorr & 0.8574  & \textbf{0.9379} \\
		\midrule
		\multicolumn{4}{c}{$m$=24} \\
		\midrule
		\multirow{2}[2]{*}{25} & TNMSE & 0.2077  & \textbf{0.1967}  \\
		& Tcorr & 0.9771  & \textbf{0.9789}  \\
		\midrule
		\multirow{2}[2]{*}{30} & TNMSE & 0.1915  & \textbf{0.1529}  \\
		& Tcorr & 0.9799  & \textbf{0.9870}  \\
		\midrule
		\multirow{2}[2]{*}{40} & TNMSE & 0.1884  & \textbf{0.1405}  \\
		& Tcorr & 0.9804  & \textbf{0.9888}  \\
		\bottomrule
	\end{tabular}%
\end{table}%

\subsection{Ablation Study}

In this subsection, we present ablation studies on dynamic channel tracking application to validate the effectiveness of each component’s strategy utilized in the proposed algorithms: $\text{PLAY}^{+}$-CS and its extension $\text{PLAY}^{+}$-CS-MMV.

\textit{Effect of {EM step}:}
In our proposed $\text{PLAY}^{+}$-CS algorithm, the weight matrix $W_t$ is automatically learned using EM algorithm, which is derived from the Partial-LSM model.
To verify the influence of {EM step}, we conducted an ablation experiment comparing two weighting schemes: (i) PLAY-CS, (ii) $\text{PLAY}^{+}$-CS.

\textit{PLAY-CS}: We use {the reconstructed signal of the first time slot} to determine the weight matrix $W_t$, i.e.,
\begin{equation}
	(W_t)_{ii} =\frac{a+1}{b+|(\hat{x}^1)_i|}, \quad \forall t\geq 1.
	\label{eq_ablation_EM_baseline}
\end{equation}

We present the quantitative results of PLAY-CS, $\text{PLAY}^{+}$-CS in Table \ref{tab:tnmse_tcorr_ablation_EM}.
From the table, it is easy to see that $\text{PLAY}^{+}$-CS performs consistently better in terms of TNMSE and TCorr.
To be specific, the priority of $\text{PLAY}^{+}$-CS over PLAY-CS validates the effectiveness of EM-based weighting scheme \eqref{eq_wii}.

\textit{Effect of {MMV}:}
In Section \ref{sec:mmv}, we proposed an extension of $\text{PLAY}^{+}$-CS algorithm with the availability of MMV. 
Thus, it is necessary to demonstrate the benefit of including multiple measurement vectors and explore how the $\text{PLAY}^{+}$-CS-MMV algorithm performs with different number of measurement vectors $P$. 
In particular, we examine the influence of $P$ on the performance of dynamic signal reconstruction over different subcarriers of CDL dataset, 
and compare the proposed baseline algorithm: $\text{PLAY}^{+}$-CS
(i.e., We independently run the $\text{PLAY}^{+}$-CS algorithm on $P$ subcarriers.)
against its extension: $\text{PLAY}^{+}$-CS-MMV.

\begin{figure*}[]
	\centering
	\subfloat[]{	\includegraphics[width=1.8in]{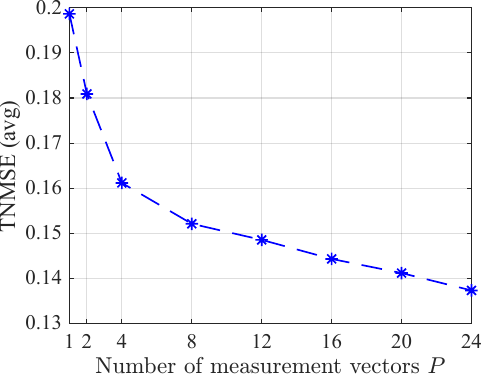}%
		\label{fig:tnmse_avg_p_snr_40}}
	\hfil
	\subfloat[]{\includegraphics[width=1.8in]{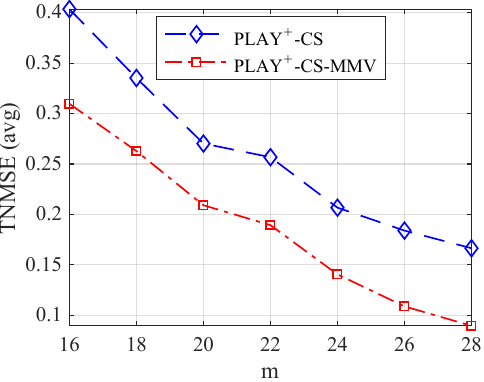}%
		\label{fig:tnmse_avg_m_snr_40}}
	\caption{(a) The TNMSE (avg) versus the number of measurement vectors $P$ when SNR of measurements is 40dB. (b) The TNMSE (avg) with respect to the measurement number $m$ when SNR is 40dB.}
	\label{fig_tnmse_avg_p_m_snr_40}
\end{figure*}

Let $m=24$, $T=200$ and SNR=40dB, we use the CDL channel dataset \cite{cdl} with $P$ subcarriers.
We run the experiment and calculate the averaged TNMSE over all subcarriers as 
$$
\text{TNMSE (avg)} := \frac{1}{TP} \sum_{t=1}^{T} \sum_{j=1}^{P} \frac{\|\hat{x}_t^{(j)} - x_t^{(j)}\|^2}{\|x_t^{(j)}\|^2}.
$$
Fig. \ref{fig:tnmse_avg_p_snr_40} shows the TNMSE (avg) of the reconstruction with respect to the number of measurement vectors $P$.
The TNMSE (avg) decreases with increasing of $P$, which demonstrates more accurate reconstruction results brought by MMV.
We further compare the baseline: $\text{PLAY}^{+}$-CS against $\text{PLAY}^{+}$-CS-MMV with $P=24$.
Fig. \ref{fig:tnmse_avg_m_snr_40} shows the TNMSE (avg) results with different measurement number $m$.
As shown in this figure, $\text{PLAY}^{+}$-CS-MMV algorithm achieves superior performance over the baseline.
This indicates that our proposed $\text{PLAY}^{+}$-CS-MMV is helpful for the performance improvement.

\section{Simulation for BSCM} \label{sec:bscm}
\subsection{Experimental Setup}

\textit{Datasets}: We applied the proposed $\text{PLAY}^{+}$-CS algorithm to the video dataset: Hall \cite{hall} \footnote{{The Hall sequence is available in http://trace.eas.asu.edu/yuv/hall\_monitor/\\hall\_qcif.7z}}.
In all experiments, we set the the first frame of the sequence as the background image, which is assumed known and unchanged.
We generated Gaussian measurement matrices $\{A_k\}_{k\geq 1}$ for all experiments.

\textit{Comparison Methods}: We compare our algorithm with the state-of-the-art in compressive background subtraction: \textbf{$\ell_1$-$\ell_1$} \cite{Mota_2016_TSP}\footnote{{Available at https://github.com/joaofcmota/AdaptiveRateCompressiveFore\\groundExtraction}} and \textbf{Modified-CS} \cite{modcs}.

\begin{figure}
	\centering
	\includegraphics[width=2.2in]{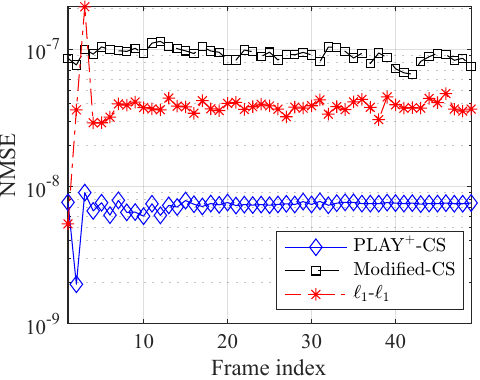}
	\caption{{The NMSE curves of different methods in the noiseless case.}}
	\label{fig:nmse_frame_index}
\end{figure}

\subsection{Performance Comparison in Noiseless Scenarios}
We performed the following experiments to validate the efficiency of our algorithm in the context of BSCM applications.
In the noiseless case of the Hall sequence, we set measurement number ${m}={2500}$.
Modified-CS \cite{modcs} reconstructs each frame using the support of the previously reconstructed frame as an aid. In our implementation, the support of a signal was computed as the set of indices containing at least 95\% of its energy.
The $\ell_1$-$\ell_1$ method \cite{Mota_2016_TSP} reconstructs each frame using prior information $\mu$. This prior information can be derived from previously recovered signals, specifically $\hat{x}_{t-1}$ and $\hat{x}_{t-2}$.
We set algorithm parameters of $\ell_1$-$\ell_1$ as in \cite{Mota_2016_TSP}: oversampling parameters $\delta_k=0.1$ for all $k$, and filter parameter $\alpha=0.5$. For motion estimation scheme used in \cite{Mota_2016_TSP}, we use the same parameters in \cite{Mota_2016_TSP}: block size $\gamma=8$, and search limit $\rho=6$.

Fig. \ref{fig:nmse_frame_index} presents the results of our experiment. 
The NMSE curves over 50 frames indicate that the reconstruction errors of $\text{PLAY}^{+}$-CS consistently remain below those of Modified-CS \cite{modcs} and $\ell_1$-$\ell_1$ \cite{Mota_2016_TSP}.
Our method not only recovers the video frames with high fidelity but also detects clear profiles of the moving foreground in greater detail. 
This demonstrates that the Partial-LSM filtering sparsity model effectively addresses the challenges posed by quick motion.

\section{Conclusion} \label{sec:conclusion}
This study addresses the challenges of dynamic channel tracking in massive MIMO systems and time-varying decomposition in video foreground-background separation.
We introduced the Partial-Laplacian model, a dynamic statistical model designed to capture the structured dynamic sparsity of real-world signal sequences.
Building on this, we developed a unified DCS framework, PLAY-CS, for dynamic signal reconstruction applications, revealing the inherent correlations among existing DCS algorithms.
Leveraging the Partial-LSM filtering sparsity model, we proposed a novel DCS algorithm, $\text{PLAY}^{+}$-CS. Additionally, by exploiting the dynamic joint sparsity in massive MIMO OFDM systems, we extended this algorithm to broadband channel reconstruction, resulting in the $\text{PLAY}^{+}$-CS-MMV algorithm. This extension was achieved through the synergistic integration of MMV technology with our developed algorithm.
The performance of $\text{PLAY}^{+}$-CS and $\text{PLAY}^{+}$-CS-MMV was validated through extensive simulations, demonstrating their effectiveness in dynamic channel tracking and online video foreground-background separation. These simulations showed that the proposed algorithms significantly outperform existing DCS algorithms by effectively exploiting the structured dynamic sparsity inherent in practical massive MIMO channels and surveillance video sequences.

\section{Acknowledgement}
This work is supported by National Key Research and Development Program of China under Grant 2021YFA1003300.

\bibliographystyle{elsarticle-num} 
\bibliography{mybib}

\begin{thebibliography}{10}
\expandafter\ifx\csname url\endcsname\relax
  \def\url#1{\texttt{#1}}\fi
\expandafter\ifx\csname urlprefix\endcsname\relax\def\urlprefix{URL }\fi
\expandafter\ifx\csname href\endcsname\relax
  \def\href#1#2{#2} \def\path#1{#1}\fi

\bibitem{mimo}
E.~G. Larsson, O.~Edfors, F.~Tufvesson, T.~L. Marzetta, Massive {MIMO} for next
  generation wireless systems, IEEE Commun. Mag. 52~(2) (2014) 186--195.

\bibitem{BSCM_2016}
W.~Cao, et~al., Total variation regularized tensor {RPCA} for background
  subtraction from compressive measurements, IEEE Trans. Image Process. 25~(9)
  (2016) 4075--4090.

\bibitem{rwl1-df}
A.~S. Charles, C.~J. Rozell, Dynamic filtering of sparse signals using
  reweighted $\ell_1$, in: Proc. Int. Conf. Acoust., Speech, Signal Process.
  (ICASSP), 2013.

\bibitem{kf}
R.~E. Kalman, A new approach to linear filtering and prediction problems, J.
  Basic Eng. 82 (1960) 35--45.

\bibitem{dcs-review}
N.~Vaswani, J.~Zhan, Recursive recovery of sparse signal sequences from
  compressive measurements: A review, IEEE Trans. Signal Process. 64~(13)
  (2016) 3523--3549.

\bibitem{batch-1}
D.~P. Wipf, B.~D. Rao, An empirical {Bayesian} strategy for solving the
  simultaneous sparse approximation problem, IEEE Trans. Signal Process. 55~(7)
  (2007) 3704--3716.

\bibitem{batch-2}
J.~A. Tropp, Algorithms for simultaneous sparse approximation-part {II}:
  {Convex} {Relaxation}, Signal Process. 86~(3) (2006) 589--602.

\bibitem{kfcs}
N.~Vaswani, Kalman filtered compressed sensing, in: Proc. IEEE Int. Conf. Image
  Process. (ICIP), 2008.

\bibitem{lscs}
N.~Vaswani, {LS}-{CS}-residual {(LS-CS)}: Compressive sensing on the least
  squares residual, IEEE Trans. Signal Process. 58~(8) (2010) 4108--4120.

\bibitem{modcs}
N.~Vaswani, W.~Lu, Modified-{CS}: Modifying compressive sensing for problems
  with partially known support, IEEE Trans. Signal Process. 58~(9) (2010)
  4595--4607.

\bibitem{regmodbpdn}
W.~Lu, N.~Vaswani, Regularized modified {BPDN} for noisy sparse reconstruction
  with partial erroneous support and signal value knowledge, IEEE Trans. Signal
  Process. 60~(1) (2012) 182--196.

\bibitem{bp}
S.~S. Chen, D.~L. Donoho, M.~A. Saunders, Atomic decomposition by basis
  pursuit, SIAM J. Sci. Comput. 20~(1) (1998) 33--61.

\bibitem{rwl1}
E.~J. Candes, M.~B. Wakin, S.~P. Boyd, Enhancing sparsity by reweighted
  $\ell_1$ minimization, J. Fourier Anal. Appl. 14~(5--6) (2008) 877--905.

\bibitem{Mota_2017_TIT}
J.~F.~C. Mota, N.~Deligiannis, M.~R.~D. Rodrigues, Compressed sensing with
  prior information: Strategies, geometry, and bounds, IEEE Trans. Inf. Theory
  63~(7) (2017) 4472--4496.

\bibitem{Mota_2016_TSP}
J.~F.~C. Mota, N.~Deligiannis, A.~C. Sankaranarayanan, V.~Cevher, M.~R.~D.
  Rodrigues, Adaptive-rate reconstruction of time-varying signals with
  application in compressive foreground extraction, IEEE Trans. Signal Process.
  64~(14) (2016) 3651--3666.

\bibitem{bp2}
E.~J. Candes, J.~K. Romberg, T.~Tao, Stable signal recovery from incomplete and
  inaccurate measurements, Commun. Pure Appl. Math. 59~(8) (2006) 1207--1223.

\bibitem{Luong_2016_ICIP}
H.~V. Luong, J.~Seiler, A.~Kaup, S.~Forchhammer, Sparse signal reconstruction
  with multiple side information using adaptive weights for multiview sources,
  in: IEEE Int. Conf. Image Process.(ICIP), 2016, pp. 2534--2538.

\bibitem{Luong_2018_TIP}
H.~V. Luong, N.~Deligiannis, J.~Seiler, S.~Forchhammer, A.~Kaup, Compressive
  online robust principal component analysis via $ n $-$\ell_1 $ minimization,
  IEEE Trans. Image Process. 27~(9) (2018) 4314--4329.

\bibitem{GSM-zhanglei}
L.~Zhang, W.~Wei, C.~Tian, F.~Li, Y.~Zhang, Exploring structured sparsity by a
  reweighted laplace prior for hyperspectral compressive sensing, IEEE Trans.
  Image Process. 25~(10) (2016) 4974--4988.

\bibitem{LSM-nips2010}
P.~Garrigues, B.~Olshausen, Group sparse coding with a laplacian scale mixture
  prior, Proc. Adv. Neural Inf. Process. Syst. (2010) 676--684.

\bibitem{cdl}
Study on channel model for frequencies from 0.5 to 100 ghz (Jun. 2018).

\bibitem{hbf-lt}
T.~Lin, J.~Cong, Y.~Zhu, J.~Zhang, K.~B. Letaief, Hybrid beamforming for
  millimeter wave systems using the mmse criterion, IEEE Trans. Commun. 67~(5)
  (2019) 3693--3708.

\bibitem{hbf-ssp}
O.~E. Ayach, S.~Rajagopal, S.~Abu-Surra, Z.~Pi, R.~W. Heath, Spatially sparse
  precoding in millimeter wave {MIMO} systems, IEEE Trans. Wireless Commun.
  13~(3) (2014) 1499--1513.

\bibitem{hall}
P.~Seeling, M.~Reisslein, Video traffic characteristics of modern encoding
  standards: H.264/{AVC} with {SVC} and {MVC} extensions and {H}.265/{HEVC},
  Scientif. World J. 2014 (2014) 1--16.

\bibitem{BSCM_2004}
M.~Piccardi, Background subtraction techniques: A review, in: Proc. IEEE Int.
  Conf. Syst., Man Cybern., Vol.~4, 2004, pp. 3099--3104.

\bibitem{BSCM_2010}
Y.~Benezeth, P.-M. Jodoin, B.~Emile, H.~Laurent, C.~Rosenberger, Comparative
  study of background subtraction algorithms, J. Electron. Imag. 19~(3) (2010)
  033003.

\bibitem{BSCM_2014}
A.~Sobral, A.~Vacavant, A comprehensive review of background subtraction
  algorithms evaluated with synthetic and real videos, Comput. Vis. Image
  Understand. 122 (2014) 4--21.

\bibitem{tipping_sbl}
M.~E. Tipping, Sparse {B}ayesian learning and the relevance vector machine, J.
  Mach. Learn. Res. 1 (2001) 211--244.

\bibitem{mmv1-l1-svd}
D.~Malioutov, A.~S. Willsky, A sparse signal reconstruction perspective for
  source localization with sensor arrays, IEEE Trans. Signal Process. 53~(8)
  (2005) 3010--3022.

\bibitem{spgl}
E.~Van Den~Berg, M.~P. Friedlander, Probing the {P}areto frontier for basis
  pursuit solutions, SIAM J. Sci. Comput. 31~(2) (2008) 890--912.

\end{thebibliography}

\end{document}